\documentclass[twocolumn, nofootinbib, nobibnotes, amsmath,amssymb,aps, pra, floatfix]{revtex4-1}

\usepackage{bbm}
\usepackage{mathtools}
\usepackage[export]{adjustbox}
\usepackage{wrapfig}
\usepackage{import}
\usepackage{bbold}
\usepackage{microtype}
\usepackage{bm} 
\usepackage{graphicx}
\usepackage{dcolumn}
\usepackage{color}
\usepackage{silence}

\renewcommand{\v}[1]{\ensuremath{\mathbf{#1}}} 
\newcommand{\gv}[1]{\ensuremath{\mbox{\boldmath$ #1 $}}} 
\newcommand{\abs}[1]{\left| #1 \right|} 
\newcommand{\pd}[2]{\frac{\partial #1}{\partial #2}} 
\newcommand{\grad}[1]{\gv{\nabla} #1} 
\renewcommand{\div}[1]{\gv{\nabla} \cdot #1} 

\begin{document}

\preprint{APS/123-QED}

\title{On the hydrodynamic canonical formalism of the Gross-Pitaevskii field}

\author{Y. Buggy}
 \author{P. {\"O}hberg}
\affiliation{SUPA, Institute of Photonics and Quantum Sciences, Heriot-Watt University, Edinburgh EH14 4AS, United Kingdom}


\begin{abstract}
We derive a canonical formalism for the hydrodynamic representation of the Gross-Pitaevskii field (nonlinear Schr\"odinger field), where the density and the phase of the condensate form a canonical pair of conjugate field variables.
To do so, we treat the meanfield as a singular Lagrangian system and apply both the Dirac-Bergmann and Faddeev-Jackiw methods.
The Faddeev-Jackiw method is found to be a more direct approach to the problem.
\\\\\\
\end{abstract}

\maketitle

\section{Introduction}
In the early twentieth century, it was discovered that fluids at low temperatures acquire unusual properties, such as flow without resistance. 
This phenomenon called superfluidity, has since attracted great interest, both from a fundamental and applications point of view. 
Superfluidity was first discovered in Helium. 
In 1995, experimentalists managed to cool and trap alkali gases \cite{anderson1995observation,davis1995bose,bradley1995evidence} to such low temperatures that quantum effects started to play a major role. 
These ultracold bosonic gases formed a Bose-Einstein condensate with exotic properties, such as superfluidity and other coherent many-body effects. 
Nowadays, Bose-Einstein condensates are used as a tool to study the properties of matter, by creating macroscopic quantum states and using these for probing effects such as quantum phase transitions and nonequilibrium effects in quantum gases. 
In contrast to superfluid $^4$He, the atomic Bose-Einstein condensate is a weakly interacting gas and therefore well described by a meanfield theory and the Gross-Pitaevskii equation \cite{pitaevskii2016bose}. 
The mechanics of this complex scalar field are analogous to those of a classical fluid \cite{Madelung1926}, where the amplitude squared $\rho$ and the phase $\theta$ of the field may be identified with the density and the velocity potential of a fluid, respectively.
This analogy is well-known and made evident most frequently through the Hamilton-Jacobi or Lagrangian formalism.
Perhaps less known, is the canonical formalism for the hydrodynamic representation of the field where $\rho$ and $\theta$ form a canonically conjugate pair.
In the condensed matter literature (see \cite{london1954superfluids,pethick2002bose,pitaevskii2016bose,zakharov1974hamiltonian,zakh1997hamiltonian,PhysRevLett.105.095302} for instance), this fact is often postulated seemingly a priori, where one observes that the correct wave equations for the field components emerge from the proposed canonical field equations.
This presupposes that $\rho$ and $\theta$ are conjugate variables to be begin with.
The conjugate nature of the hydrodynamic variables has been shown to result from general symmetry properties of the system \cite{pokrovskii1976hamiltonian,khalatnikov1978canonical,khalatnikov1991hydrodynamics}, where $-\rho$ and $\theta$ play the role of the momentum and coordinate field variables, respectively.
In this paper, we examine an alternative forward derivation of the problem, by treating the Gross-Pitaevskii field as a singular Lagrangian system and applying both the Dirac-Bergmann and Faddeev-Jackiw methods.
\section{The Gross-Pitaevskii field}
The Hamiltonian of a dilute cloud of $N$ Bose atoms with weak contact interactions, may be written as
\begin{equation}
  \hat{H}=\sum_{i}\left( \frac{\hat{\v{p}}_i^2}{2m}+V\left( \v{r}_i \right) \right)+\sum_{i\neq j}g\delta\left( \v{r}_i-\v{r}_j \right),
  \label{eq:HamiltonianGPMicroscopic}
\end{equation}
where $\hat{\v{p}}_i=-i\hbar\grad{}_{\v{r}_i}$, $g=4\pi\hbar^2a/m$, $a$ is the $s$-wave scattering length and $V$ is an external potential.
The Gross-Pitaevskii equation may be retrieved as a meanfield approximation to the true dynamics of the system, by treating the many-body wavefunction as a product of identical single particle states:
\begin{equation}
  \Psi\left( \v{r}_1,\v{r}_2,\cdots,\v{r}_N \right)=\prod_{i=1}^N\phi\left( \v{r}_i \right),
  \label{eq:manyBodyWavefunctionProductSeperableStates}
\end{equation}
where $\int d^3\v{r}\abs{\phi\left( \v{r} \right)}^2 = 1$.
To do so, it is convenient to introduce the Lagrangian of the many-body system \cite{frenkel1934wave,dirac1981principles,schiff1968quantum,gray2004progress,deumens1994time}
\begin{equation}
  L=\int \prod_{i=1}^Nd^3\v{r}_i\Psi^*\left( i\hbar\partial_t-\hat{H} \right)\Psi,
  \label{eq:LagrangianManyBodySystemVariationalApproach}
\end{equation}
whose variation with respect to $\Psi^*$ yields the many-body Schr\"odinger equation.
Inserting Eq. (\ref{eq:manyBodyWavefunctionProductSeperableStates}) for $\Psi$ into Eq. (\ref{eq:LagrangianManyBodySystemVariationalApproach}), leads to the meanfield Lagrangian
\begin{equation}
  L_{MF}=\int d^3\v{r}\psi^*\left( i\hbar\partial_t-\hat{H}_{MF} \right)\psi,
  \label{eq:meanfieldLagrangianGPfield}
\end{equation}
where the meanfield Hamiltonian, $\hat{H}_{MF}$, acts on the macroscopic wavefunction, $\psi\left( \v{r} \right)=\sqrt{N}\phi\left( \v{r} \right)$, as
\begin{equation}
  \hat{H}_{MF}\psi=\left(-\frac{\hbar^2}{2m}\nabla^2+V+\frac{g}{2}\abs{\psi}^2\right)\psi.
  \label{eq:meanfieldHamiltonianGPfieldActionOnWavefunction}
\end{equation}
Since henceforth our concern lies only with the meanfield, we drop the subscript $MF$.
In accordance with Eqs. (\ref{eq:meanfieldLagrangianGPfield}) and (\ref{eq:meanfieldHamiltonianGPfieldActionOnWavefunction}), the meanfield Lagrangian density, may be presented, as
\begin{equation}
  \mathcal{L} = \frac{i\hbar}{2}\left(\psi^*\dot{\psi}-\psi\dot{\psi}^*\right) -\frac{\hbar^2}{2m}\grad{\psi}\cdot\grad{\psi}^* -\frac{g}{2}\abs{\psi}^4-V\abs{\psi}^2,
\label{eq:Lagrangian_GP_Psi}
\end{equation}
where we have denoted partial differentiation with respect to time by a dot and performed the transformation $\mathcal{L}\rightarrow\mathcal{L}-\frac{i\hbar}{2}\partial_t\left( \psi^*\psi \right)$, such that $\mathcal{L}\in{\rm I\!R}$.
Inserting the Lagrangian density (\ref{eq:Lagrangian_GP_Psi}) into the Euler-Lagrange field equation for $\psi^*$,
\begin{equation}
  \frac{\delta L}{\delta\psi^*}-\pd{}{t}\left(\frac{\delta L}{\delta\dot{\psi^*}}\right)=0,
  \label{}
\end{equation}
yields the Gross-Pitaevskii equation 
\begin{equation}
  i\hbar\partial_t\psi=\left( -\frac{\hbar^2}{2m}\grad{}^2+V+g\abs{\psi}^2 \right)\psi,
  \label{eq:GPequation}
\end{equation}
where $\delta L/\delta\psi=\partial\mathcal{L}/\partial\psi-\grad{}\cdot\left( \partial\mathcal{L}/\partial\left( \grad{}\psi \right) \right)$ and $\delta L/\delta\dot{\psi}=\partial\mathcal{L}/\partial\dot{\psi}$ are functional derivatives \cite{mercier2004analytical,cohen1992atom}.
Carrying out the same procedure for $\psi$, yields the complex conjugate of the Gross-Pitaevskii equation.
\section{\label{sec:GPfieldSingularLagrangianSystem}The Gross-Pitaevskii field as a singular Lagrangian system}
Notice how a complex field variable, $\psi$, automatically requires that $\mathcal{L}$ also depend on $\psi^*$ in order for the action to be real \cite{cohen1992atom}. 
Yet, the fact that the Gross-Pitaevskii equation is first order in time, signifies that $\psi$, $\psi^*$, $\dot{\psi}$ and $\dot{\psi}^*$ are not independent and that an excess of dynamical variables are contained in the Lagrangian \cite{cohen1992atom,schiff1968quantum}. 
This is invariably the situation when the Lagrangian density is linear in the time derivatives of the fields.
To see this, let $\mathcal{L}\left(\phi_{\alpha},\grad{\phi_{\alpha}}, \dot{\phi}_{\alpha}\right)$ be a multi-component Lagrangian density which is linear in the time derivatives of the fields, where $\phi_{\alpha}\equiv\phi_1,\cdots,\phi_n$.
In this situation, the total Lagrangian of the system, may be written as
\begin{equation}
  L\left[ \phi_{\alpha},\dot{\phi}_{\alpha} \right]=\int d^3\v{r}\sum_i\mathcal{A}_i\left(\phi_{\alpha},\grad{\phi}_{\alpha}\right)\dot{\phi}_i-U \left[\phi_{\alpha}\right],
  \label{eq:Lagrangian_density_first_order_in_velocities}
\end{equation}
where $U$ is an interaction functional of the field components:
\begin{equation}
  U\left[\phi_{\alpha}\right]=\int d^3\v{r} \mathcal{U}\left(\phi_{\alpha},\grad{\phi}_{\alpha}\right).
  \label{eq:interaction_functional}
\end{equation}
When $L$ is of the form (\ref{eq:Lagrangian_density_first_order_in_velocities}), the canonical momenta, $\pi_i=\mathcal{A}_i\left( \phi_{\alpha},\grad{}\phi_{\alpha} \right)$, can not be treated as independent dynamical variables.  
Indeed, upon constructing the total Hamiltonian 
\begin{equation}
  H\left[ \phi_{\alpha},\pi_{\alpha} \right]=\int d^3\v{r}\sum_{i=1}^n \pi_i\dot{\phi}_i-L\left[ \phi_{\alpha},\dot{\phi}_{\alpha} \right],
  \label{eq:legendreTransformFields}
\end{equation}
the field velocities $\dot{\phi}_i$ no longer appear on the right hand side of Eq. (\ref{eq:legendreTransformFields}) as they would typically.
Hence, the Hamiltonian reduces to the interaction functional, $H=U$, and it is not possible to invert $\dot{\phi}_i$ as a function of $\phi_{\alpha}$ and $\pi_{\alpha}$, since 
\begin{equation}
  \frac{\partial^2\mathcal{L}}{\partial\dot{\phi}_j\partial\dot{\phi}_i}=\pd{\mathcal{A}_i}{\dot{\phi}_j}=0.
  \label{eq:non_invertibility_condition}
\end{equation}
Dynamical systems with this property are called \textit{singular Lagrangian systems} or \textit{constrained Hamiltonian systems} \cite{henneaux1992quantization,sundermeyer1982constrained}.
Two equivalent \cite{barcelos1992symplectic} methods have been devised to eliminate redundant variables and construct a reduced phase space for such systems: the \textit{Dirac-Bergmann} method \cite{dirac1958generalized,dirac2001lectures,sundermeyer1982constrained,das2008lectures,gitman2012quantization,rothe2010classical,burnel2009canonical,mitra2014symmetries,maskawa1976singular,weinberg1995quantum} and the \textit{Faddeev-Jackiw} method \cite{faddeev1988hamiltonian,jackiw1994quantization,drummond2014quantum,ma14ller2006introduction,rothe2010classical}.
In the particular case of the Schr{\"o}dinger field, an alternative route is made available by performing a suitable canonical transformation \cite{kerman1976hamiltonian,cohen1992atom}, where one begins by decomposing the field into real and imaginary parts and then supplements the resulting Lagrangian by a total time derivative to obtain a single pair of real conjugate variables.
Several other field transformations yielding Schr{\"o}dinger's equation from a canonical field equation, can also be found in the literature \cite{henley1959elementary,schiff1968quantum}.
However, these depend either on one or two complex pairs of conjugate variables, meaning redundant variables have not entirely been eliminated.
For a review of these formalisms, and an application of the Dirac-Bergmann and the Faddeev-Jackiw methods to the Schr{\"o}dinger field, see \cite{gergely2002hamiltonian}.\\
In this paper, we take a different approach and derive a canonical formalism in terms of the single pair of real variables $\left(\rho,\theta\right)$, namely the density and the phase of the matter-field $\psi$.
These constitute the natural pair of variables connecting the field and fluid descriptions of a condensate.
Accordingly, we shall refer to this formalism as the ``hydrodynamic canonical formalism''.
\section{\label{sec:canonical_formalism}Hydrodynamic formalism for the Gross-Pitaevskii field}
Following up on previous discussions, let us represent the Gross-Pitaevskii field in the polar, or Madelung form
\begin{equation}
  \psi=\sqrt{\rho}e^{i\frac{\theta}{\hbar}},\quad\quad\psi^*=\sqrt{\rho}e^{-i\frac{\theta}{\hbar}},
\label{eq:polar_form_psi}
\end{equation}
and treat $\left( \rho,\theta \right)$ as the independent variables subject to the process of variation.
From a dynamical perspective, a clear role is played by the new field variables: the first defines the distribution or amplitude of the matter-field, while the second dictates the flow of this distribution.
This may be seen by substituting Eq. (\ref{eq:polar_form_psi}) into the current density
\begin{equation}
\v{J}=\frac{i\hbar}{2m}\left(\psi\grad{\psi^*}-\psi^*\grad{\psi}\right),
\label{eq:current_density}
\end{equation}
which yields the more perspicuous form
\begin{equation}
\v{J}=\frac{\rho}{m}\grad{\theta},
\label{eq:current_density_polar_form}
\end{equation}
and allows for the identification of the velocity field
\begin{equation}
\v{v}=\frac{\grad{\theta}}{m}.
\label{eq:velocity_field_polar_form}
\end{equation}
\subsection{\label{sec:GPfieldHydrodynamicLagrangianFormalism}Lagrangian formalism}
Under substitution (\ref{eq:polar_form_psi}), the Lagrangian density (\ref{eq:Lagrangian_GP_Psi}), becomes
\begin{equation}
  \mathcal{L}=-\rho\left(\dot{\theta}+\frac{(\grad{\theta})^2}{2m}+\frac{g}{2}\rho+V\right) -\frac{\hbar^2}{8m\rho}(\grad{\rho})^2.
\label{eq:Lagrangian_GP_polar}
\end{equation}
The Euler-Lagrange field equations for $\theta$ and $\rho$, yield respectively, the wave equations
\begin{align}
  &\partial_t\rho+\div{\v{J}}=0, \label{eq:continuityGPfield} \\
  &\partial_t\theta+\frac{1}{2}mv^2+g\rho+V+Q=0, \label{eq:QHJEGPfield}
\end{align}
where $v=\abs{\v{v}}$ is the modulus of the superfluid flow from Eq. (\ref{eq:velocity_field_polar_form}), $\v{J}$ is given by Eq. (\ref{eq:current_density_polar_form}) and
\begin{equation}
  Q=-\frac{\hbar^2}{2m}\frac{\nabla^2\sqrt{\rho}}{\sqrt{\rho}},
  \label{eq:quantum_potential}
\end{equation}
is the \textit{quantum potential} \cite{bohm1952suggestedI}.
The coupled wave equations (\ref{eq:continuityGPfield}) and (\ref{eq:QHJEGPfield}), express, respectively, the conservation of mass and momentum.
They are entirely equivalent to the Gross-Pitaevskii Eq. (\ref{eq:GPequation}), where transformation (\ref{eq:polar_form_psi}) defines the mapping between both sets of equations.
\subsection{\label{sec:GPfieldHydrodynamicCanonicalFormalism}Canonical formalism}
Let us turn our attention to the construction of a canonical formalism for the field.
In particular, we will show that the reduced phase space of the system comprises the single pair of conjugate variables $\left( \rho,\theta \right)$, dynamically governed by the following canonical field equations:
\begin{align}
  \dot{\rho}\left( \v{r} \right)&=\frac{\delta H}{\delta\theta\left( \v{r} \right)}, \label{eq:canonicalHydrodynamicalFieldEquationRhoDot}\\
  \dot{\theta}\left( \v{r} \right)&=-\frac{\delta H}{\delta\rho\left( \v{r} \right)}, \label{eq:canonicalHydrodynamicalFieldEquationThetaDot}
\end{align}
where $H$ is the total canonical Hamiltonian of the field, introduced further in the text.
In turn, the Poisson bracket of two dynamical variables $f$ and $g$ on the reduced phase space, takes the form
\begin{equation}
  \left\{ f\left( \v{x} \right),g\left( \v{y} \right) \right\}=\int d^3\v{r}\left( \frac{\delta f\left( \v{x} \right)}{\delta\rho\left( \v{r} \right)}\frac{\delta g\left( \v{y} \right)}{\delta\theta\left( \v{r} \right)}-\frac{\delta f\left( \v{x} \right)}{\delta\theta\left( \v{r} \right)}\frac{\delta g\left( \v{y} \right)}{\delta\rho\left( \v{r} \right)} \right).
  \label{eq:poissonBracketReducedPhaseSpace}
\end{equation}
In the following section, we retrieve the above expression for the Poisson bracket as the reduced Dirac bracket on the full phase.
\subsubsection{\label{sec:GPfieldHydrodynamicCanonicalFormalismDBmethod}The Dirac-Bergmann algorithm}
As a starting point to applying the Dirac-Bergmann algorithm, let us transform the Lagrangian density (\ref{eq:Lagrangian_GP_polar}), according to
\begin{align}
  \mathcal{L}&\rightarrow\mathcal{L}+\pd{}{t}\left( \frac{\rho\theta}{2} \right) \label{eq:LagrangianDensityTransformationTotalTimeDerivative} \\
  &=\frac{1}{2}\left( \theta\dot{\rho}-\rho\dot{\theta} \right)-\rho\left( \frac{\left( \grad{\theta} \right)^2}{2m}+\frac{g}{2}\rho+V \right)-\frac{\hbar^2}{8m\rho}\left( \grad{\rho} \right)^2\nonumber.
\end{align}
This will allow for the possibility of readily identifying a canonical transformation (\ref{eq:canonicalTransformationHydrodynamicalFieldReduction}), which separates out the physical pair of conjugate variables from the redundant pair of conjugate variables, the latter representing the constraints of the theory.
The first order nature of the Lagrangian density (\ref{eq:LagrangianDensityTransformationTotalTimeDerivative}), means that the canonical momenta
\begin{equation}
  \pi_{\rho}=\pd{\mathcal{L}}{\dot{\rho}}=\frac{\theta}{2},\quad\quad\quad\pi_{\theta}=\pd{\mathcal{L}}{\dot{\theta}}=-\frac{\rho}{2},
  \label{eq:conjugate_momenta_2}
\end{equation}
cannot be treated as independent dynamical variables.
In other words, the phase space variables are restricted by the following equations:
\begin{equation}
  C_1=\pi_{\theta}+\frac{\rho}{2}=0, \quad\quad C_2=\pi_{\rho}-\frac{\theta}{2}=0.
  \label{eq:constraintEquationsPhaseSpaceHydrodynamicCanonicalFormalism}
\end{equation}
In the Dirac treatment \cite{dirac1958generalized,dirac2001lectures} of singular Lagrangian systems, the relations from Eq. (\ref{eq:constraintEquationsPhaseSpaceHydrodynamicCanonicalFormalism}) define a constraint hypersurface $\Gamma_c$ in the full phase space $\left( \rho,\theta,\pi_{\rho},\pi_{\theta} \right)$.
Since the constraints are primary, only two canonical variables play a physical role in the dynamical description of the system.\\
Before constructing the reduced phase space, it is instructive to examine the form of the dynamical equations on $\Gamma_c$, as embedded in the full phase space of the system.
Although the duality of the Legendre transform is destroyed by the singular nature of the system, let us define a canonical Hamiltonian density, $\mathcal{H}$, according to the usual prescription
\begin{equation}
  \mathcal{H}=\pi_{\rho}\dot{\rho}+\pi_{\theta}\dot{\theta}-\mathcal{L},
  \label{}
\end{equation}
which on $\Gamma_c$, is given by
\begin{equation}
  \mathcal{H}\approx\rho\left[ \frac{(\grad{\theta})^2}{2m}+\frac{g}{2}\rho+V\right]+\frac{\hbar^2}{8m\rho}\left(\grad{\rho}\right)^2,
  \label{eq:Hamiltonian_density_GP}
\end{equation}
where the symbol '$\approx$' denotes weak equality \cite{dirac2001lectures}.
The primary Hamiltonian \cite{sundermeyer1982constrained}, which incorporates the primary constraints, may be defined as
\begin{equation}
  H_p=\int d^3\v{r}\left( \mathcal{H}+\dot{\theta}C_1+\dot{\rho}C_2 \right),
  \label{eq:hamiltonianPrimaryGPfield}
\end{equation}
where the Lagrange multipliers are identified \cite{das2008lectures} with the field velocities $\dot{\theta}$ and $\dot{\rho}$, which are unknown functions on phase space.
Note also that $\mathcal{H}_p\approx\mathcal{H}$.
On the full phase space, the time evolution of an arbitrary dynamical variable $f$ is generated by the primary Hamiltonian rather than the canonical Hamiltonian \cite{dirac2001lectures,gitman2012quantization,rothe2010classical,das2008lectures}, through the Poisson bracket
\begin{equation}
  \dot{f}\approx\left\{ f,H_p \right\},
  \label{eq:timeDerivativePoissonBracketWeakEquality}
\end{equation}
where $\left\{ f,g \right\}\equiv \left\{ f\left( \v{x} \right),g\left( \v{y} \right) \right\}$. 
Note that we will occasionally use the shorthand form $\left\{ f,g \right\}=\left\{ f,g \right\}_{\rho,\pi_{\rho}}+\left\{ f,g \right\}_{\theta,\pi_{\theta}}$.
The unknown functions $\dot{\rho}$ and $\dot{\theta}$, may be obtained on $\Gamma_c$, from the consistency requirement that the constraint equations must be preserved in time: $\dot{C}_i\approx0$.
To see this, let us replace $f$ in Eq. (\ref{eq:timeDerivativePoissonBracketWeakEquality}) by $C_i$ and expand out the Poisson bracket, which gives
\begin{multline}
  \dot{ C}_i\left( \v{r} \right)\approx \int d^3\v{r}'\left( \frac{\delta C_i\left( \v{r} \right)}{\delta\rho\left( \v{r}' \right)}\frac{\delta H_p}{\delta\pi_{\rho}\left( \v{r}' \right)}- \frac{\delta C_i\left( \v{r} \right)}{ \delta\pi_{\rho}\left( \v{r}' \right)}\frac{\delta H_p}{ \delta\rho\left( \v{r}' \right)} \right. \\
  \left. +\frac{\delta C_i\left( \v{r} \right)}{\delta\theta\left( \v{r}' \right)}\frac{\delta H_p}{\delta\pi_{\theta}\left( \v{r}' \right)}- \frac{\delta C_i\left( \v{r} \right)}{ \delta\pi_{\theta}\left( \v{r}' \right)}\frac{\delta H_p}{ \delta\theta\left( \v{r}' \right)}\vphantom{\int}\right).
  \label{eq:timeDerivativeConstraintsExpandedPoissonBracket}
\end{multline}
For the constraint $C_1=\pi_{\theta}+\rho/2$, the first and last of the four terms under the integral survive, yielding
\begin{equation}
  \dot{C}_1\left( \v{r} \right)\approx\frac{1}{2}\frac{\delta H_p}{\delta\pi_{\rho}(\v{r})}-\frac{\delta H_p}{\delta\theta\left( \v{r} \right)},
  \label{eq:timeDerivativeConstraint1}
\end{equation}
while for $C_2=\pi_{\rho}-\theta/2$, the other two terms survive, yielding
\begin{equation}
  \dot{C}_2\left( \v{r} \right)\approx-\frac{\delta H_p}{\delta\rho\left( \v{r} \right)}-\frac{1}{2}\frac{\delta H_p}{\delta\pi_{\theta}\left( \v{r} \right)}.
  \label{eq:timeDerivativeConstraint2}
\end{equation}
Using Eq. (\ref{eq:hamiltonianPrimaryGPfield}) for $H_p$, with $C_i$ given by Eq. (\ref{eq:constraintEquationsPhaseSpaceHydrodynamicCanonicalFormalism}), we then find
\begin{align}
  \dot{C}_1&\approx\frac{\dot{\rho}}{2}-\frac{\delta H}{\delta\theta}+\frac{\dot{\rho}}{2}\approx0, \label{eq:timeDerivativeConstraintCanonicalFieldEquation1}\\
  \dot{C}_2&\approx-\frac{\delta H}{\delta\rho}-\frac{\dot{\theta}}{2}-\frac{\dot{\theta}}{2}\approx0.
  \label{eq:timeDerivativeConstraintCanonicalFieldEquation2}
\end{align}
Hence the consistency equations for the constraints are equivalent to the canonical system of equations (\ref{eq:canonicalHydrodynamicalFieldEquationRhoDot}) and (\ref{eq:canonicalHydrodynamicalFieldEquationThetaDot}).
However, these have emerged as weak equalities on $\Gamma_c$, where substitution of the canonical Hamiltonian density (\ref{eq:Hamiltonian_density_GP}) into Eqs. (\ref{eq:timeDerivativeConstraintCanonicalFieldEquation1}) and (\ref{eq:timeDerivativeConstraintCanonicalFieldEquation2}), yields, respectively, Eqs. (\ref{eq:continuityGPfield}) and (\ref{eq:QHJEGPfield}), also in the form of weak equalities.\\\\
Let us now proceed with the phase space reduction.
This may be achieved by appropriate implementation of Dirac brackets instead of Poisson brackets. 
The Dirac bracket of two phase space variables, is given by \cite{dirac1958generalized,dirac2001lectures,sundermeyer1982constrained,das2008lectures,burnel2009canonical,rothe2010classical,gitman2012quantization,mitra2014symmetries}
\begin{equation}
  \left\{ f\left( \v{x}\right),g\left( \v{y}  \right) \right\}_D=\left\{ f\left( \v{x} \right),g\left( \v{y} \right) \right\} -\sum_{i,j}R_{ij},  \label{eq:diracBracketDefinitionFields}
\end{equation}
where we have defined
\begin{equation}
  R_{ij}=\iint d^3\v{r}d^3\v{r}'\left\{ f\left( \v{x} \right), C_i\left( \v{r} \right)\right\}Q_{ij}^{-1}\left( \v{r},\v{r}' \right)\left\{ C_j\left( \v{r}' \right),g\left( \v{y} \right) \right\},
  \label{eq:DiracBracketReductionContributionRij}
\end{equation}
and $Q_{ij}\left( \v{r},\v{r}' \right)=\left\{ C_i\left( \v{r}\right),C_j\left( \v{r}' \right) \right\}$ is a matrix with elements given by the Poisson brackets of the constraints, which satisfies
\begin{equation}
  \sum_j\int d^3\v{r}'' Q_{ij}\left( \v{r},\v{r}'' \right)Q^{-1}_{jk}\left( \v{r}'',\v{r}' \right)=\delta_{ik}\delta\left( \v{r}-\v{r}' \right).
  \label{eq:PoissonBracketConstrainMatrixInverseIdensityRelationFields}
\end{equation}
Furthermore, in contrast to Eq. (\ref{eq:timeDerivativePoissonBracketWeakEquality}), time-evolution is now generated by the canonical Hamiltonian, through the Dirac bracket
\begin{equation}
  \dot{f}=\left\{ f,H \right\}_D.
  \label{eq:timeDerivativeDiracBracket}
\end{equation}
Recalling Eq. (\ref{eq:constraintEquationsPhaseSpaceHydrodynamicCanonicalFormalism}) for the constraints, let us explicitly construct $Q_{ij}\left( \v{r},\v{r}' \right)$ for our system.
In view of the anti-symmetric property of the Poisson bracket, the diagonal elements are $Q_{11}=\left\{ C_1\left( \v{r} \right),C_1 \left( \v{r}' \right)\right\}=Q_{22}=\left\{ C_2\left( \v{r} \right),C_2\left( \v{r}' \right) \right\}=0$.
For the off-diagonal elements, we find
\begin{align}
  Q_{12}&=\left\{ C_1\left( \v{r} \right),C_2\left( \v{r}' \right) \right\}_{\rho,\pi_{\rho}} +\left\{ C_1\left( \v{r}\right) ,C_2\left( \v{r}' \right) \right\}_{\theta,\pi_{\theta}}\nonumber \\
  &=\delta\left( \v{r}-\v{r}' \right)=-Q_{21}.
  \label{}
\end{align}
Thus, the constraint Poisson bracket matrix, takes the form
\begin{equation}
  Q\left( \v{r},\v{r}' \right)=
  \begin{pmatrix}
   0&1\\
   -1&0
  \end{pmatrix}
  \delta\left( \v{r}-\v{r}' \right),
  \label{eq:poissonBracketConstraintMatrix}
\end{equation}
whose inverse is given by
\begin{equation}
  Q^{-1}\left( \v{r},\v{r}' \right)=
  \begin{pmatrix}
   0&-1\\
   1&0
  \end{pmatrix}
  \delta\left( \v{r}-\v{r}' \right),
  \label{eq:poissonBracketConstraintMatrixInverse}
\end{equation}
in accordance with Eq. (\ref{eq:PoissonBracketConstrainMatrixInverseIdensityRelationFields}).
For the present problem, evaluation of the Dirac bracket (\ref{eq:diracBracketDefinitionFields}) involves the two off-diagonal terms $R_{12}$ and $R_{21}$.
Since both constraints (\ref{eq:constraintEquationsPhaseSpaceHydrodynamicCanonicalFormalism}) depend on two phase space variables, only two out of four terms are retained in any Poisson bracket comprised in $R_{12}$ and $R_{21}$.
Substituting Eqs. (\ref{eq:constraintEquationsPhaseSpaceHydrodynamicCanonicalFormalism}) and (\ref{eq:poissonBracketConstraintMatrixInverse}) into Eq. (\ref{eq:DiracBracketReductionContributionRij}) and making use of standard delta function relations, gives
\begin{equation}
  R_{12}=\int d^3\v{r}\left( -\frac{1}{2}\frac{\delta f\left( \v{x} \right)}{\delta\pi_{\rho}\left( \v{r} \right)}+\frac{\delta f\left( \v{x} \right)}{\delta\theta\left( \v{r} \right)} \right)\left( \frac{\delta g\left( \v{y} \right)}{\delta\rho\left( \v{r} \right)}+\frac{1}{2}\frac{\delta g\left( \v{y} \right)}{\delta\pi_{\theta}\left( \v{r} \right)} \right),
  \label{eq:RoneTwo}
\end{equation}
while pursuing a similar procedure for $R_{21}$, leads to 
\begin{equation}
  R_{21}=\int d^3\v{r}\left( \frac{\delta f\left( \v{x} \right)}{\delta\rho\left( \v{r} \right)}+\frac{1}{2}\frac{\delta f\left( \v{x} \right)}{\delta\pi_{\theta}\left( \v{r} \right)} \right)\left( \frac{1}{2}\frac{\delta g\left( \v{y} \right)}{\delta\pi_{\rho}\left( \v{r} \right)}-\frac{\delta g\left( \v{y} \right)}{\delta\theta\left( \v{r} \right)} \right).
  \label{eq:RTwoOne}
\end{equation}
Hence, from Eqs. (\ref{eq:diracBracketDefinitionFields}), (\ref{eq:RoneTwo}) and (\ref{eq:RTwoOne}), we find that the Dirac bracket takes the form
\begin{multline}
  \left\{ f,g \right\}_D=\left\{ f,g \right\}_{\rho,\theta}+\frac{1}{2}\left[ \left\{ f,g \right\}_{\rho,\pi_{\rho}}+\left\{ f,g \right\}_{\theta,\pi_{\theta}} \right]\\+\frac{1}{4}\left\{ f,g \right\}_{\pi_{\rho},\pi_{\theta}},
  \label{eq:diracBracketHydrodynamicalFieldFullPhaseSpace}
\end{multline}
where we have used the antisymmetry of brackets under exchange of field variables, e.g. $\left\{ f,g \right\}_{\rho,\theta}=-\left\{ f,g \right\}_{\theta,\rho}$.
From here, the Dirac brackets of the canonical field variables may be read off, as
\begin{equation}
  \left\{ \rho\left( \v{r} \right),\rho\left( \v{r}' \right) \right\}_D=\left\{ \theta\left( \v{r} \right),\theta\left( \v{r}' \right) \right\}_D=0,
  \label{}
\end{equation}
\begin{equation}
  \left\{ \pi_{\rho}\left( \v{r} \right),\pi_{\rho}\left( \v{r}' \right) \right\}_D=\left\{ \pi_{\theta}\left( \v{r} \right),\pi_{\theta}\left( \v{r}' \right) \right\}_D=0
  \label{}
\end{equation}
\begin{equation}
  \left\{ \rho\left( \v{r} \right),\pi_{\theta}\left( \v{r}' \right) \right\}_D=\left\{ \theta\left( \v{r} \right),\pi_{\rho}\left( \v{r}' \right) \right\}_D=0,
  \label{}
\end{equation}
\begin{equation}
  \left\{ \rho\left( \v{r} \right),\pi_{\rho}\left( \v{r}' \right) \right\}_D=\left\{ \theta\left( \v{r} \right),\pi_{\theta}\left( \v{r}' \right) \right\}_D=\frac{1}{2}\delta\left( \v{r}-\v{r}' \right),
  \label{}
\end{equation}
\begin{equation}
  \left\{ \pi_{\rho}\left( \v{r} \right),\pi_{\theta}\left( \v{r}' \right) \right\}_D=\frac{1}{4}\delta\left( \v{r}-\v{r}' \right),
  \label{}
\end{equation}
\begin{equation}
  \left\{ \rho\left( \v{r} \right),\theta\left( \v{r}' \right) \right\}_D=\delta\left( \v{r}-\v{r}' \right).
  \label{}
\end{equation}
The reduction of phase space to the physical degrees of freedom of the system may be achieved by a suitable canonical transformation.
In particular, we would like to find a transformation
\begin{equation}
  \begin{pmatrix}
   \rho&\theta\\ 
   \pi_{\rho}&\pi_{\theta}
  \end{pmatrix}
  \quad\rightarrow\quad
  \begin{pmatrix}
    Q_1 & Q_2 \\
    P_1 & P_2
  \end{pmatrix},
  \label{}
\end{equation}
which transforms one of the set of conjugate variables, $\left( \theta,\pi_{\theta} \right)$ say, into the pair of constraints (\ref{eq:constraintEquationsPhaseSpaceHydrodynamicCanonicalFormalism}), such that $Q_2=\pi_{\theta}+\rho/2$ and $P_2=\pi_{\rho}-\theta/2$.
The existence of such a transformation is guaranteed by a theorem due to Maskawa and Nakajima \cite{maskawa1976singular}.
Under the transformation, the Dirac bracket on the full phase space should reduce to the Poisson bracket on the reduced phase space \cite{weinberg1995quantum}, so that $\left\{ f,g \right\}_D=\left\{ f,g \right\}_{Q_1,P_1}$.
In order for the transformation to be canonical, the Poisson brackets of the new variables with respect to the old variables, should satisfy
\begin{align}
  \left\{ Q_i\left( \v{r} \right),Q_j\left( \v{r}' \right) \right\}&=\left\{ P_i\left( \v{r} \right),P_j\left( \v{r}' \right) \right\}=0, \nonumber \\
  \left\{ Q_i\left( \v{r} \right),P_j\left( \v{r}' \right)\right\}&=\delta_{ij}\delta\left( \v{r}-\v{r}' \right).
  \label{eq:canonicalTransformationFieldConditions}
\end{align}
Given these observations, it is not difficult to see that the transformation
\begin{equation}
  \begin{pmatrix}
   \rho&\theta\\ 
   \pi_{\rho}&\pi_{\theta}
  \end{pmatrix}
  \quad\rightarrow\quad
  \begin{pmatrix}
    Q_1=\rho/2-\pi_{\theta}& Q_2=\pi_{\theta}+\rho/2 \\
    P_1=\theta/2+\pi_{\rho}& P_2=\pi_{\rho}-\theta/2
  \end{pmatrix},
  \label{eq:canonicalTransformationHydrodynamicalFieldReduction}
\end{equation}
transforms the second pair of conjugate variables into the constraints and is canonical.
Let us examine the form of the Dirac bracket (\ref{eq:diracBracketHydrodynamicalFieldFullPhaseSpace}) written in terms of the new canonical variables.
For this purpose, it is useful to express the functional derivatives with respect to the old variables, in terms of the new ones.
Using simple chain rules, we find
\begin{align}
    &\frac{\delta}{\delta\rho}=\frac{1}{2}\left( \frac{\delta}{\delta Q_1}+\frac{\delta}{\delta Q_2} \right),\quad\quad\frac{\delta}{\delta\theta}=\frac{1}{2}\left( \frac{\delta}{\delta P_1}-\frac{\delta}{\delta P_2} \right), \nonumber \\
    &\frac{\delta}{\delta\pi_{\rho}}=\frac{\delta}{\delta P_1}+\frac{\delta}{\delta P_2}, \quad\quad\quad\quad\frac{\delta}{\delta\pi_{\theta}}=-\frac{\delta}{\delta Q_1}+\frac{\delta}{\delta Q_2}.\label{eq:functionalDerivativeTransformationsCanonicalTransformationHydrodynamicalPhaseSpaceReduction}
\end{align}
From the sign of each of these terms and the ordering of the fields in the brackets appearing in the Dirac bracket (\ref{eq:diracBracketHydrodynamicalFieldFullPhaseSpace}), each bracket contribution may be read off, as
\begin{widetext}
\begin{align}
  \left\{ f,g \right\}_{\rho,\theta}&=\frac{1}{4}\left[ \left\{ f,g \right\}_{Q_1,P_1}-\left\{ f,g \right\}_{Q_1,P_2}+\left\{ f,g \right\}_{Q_2,P_1}-\left\{ f,g \right\}_{Q_2,P_2} \right],\\
  \frac{1}{2}\left\{ f,g \right\}_{\rho,\pi_{\rho}}&=\frac{1}{4}\left[ \left\{ f,g \right\}_{Q_1,P_1}+\left\{ f,g \right\}_{Q_1,P_2}+\left\{ f,g \right\}_{Q_2,P_1}+\left\{ f,g \right\}_{Q_2,P_2} \right], \\
  \frac{1}{2}\left\{ f,g \right\}_{\theta,\pi_{\theta}}&=\frac{1}{4}\left[ \left\{ f,g \right\}_{Q_1,P_1}-\left\{ f,g \right\}_{Q_1,P_2}-\left\{ f,g \right\}_{Q_2,P_1}+\left\{ f,g \right\}_{Q_2,P_2} \right], \\
  \frac{1}{4}\left\{ f,g \right\}_{\pi_{\rho},\pi_{\theta}}&=\frac{1}{4}\left[ \left\{ f,g \right\}_{Q_1,P_1}+\left\{ f,g \right\}_{Q_1,P_2}-\left\{ f,g \right\}_{Q_2,P_1}-\left\{ f,g \right\}_{Q_2,P_2} \right].
  \label{}
\end{align}
\end{widetext}
Therefore, the Dirac bracket (\ref{eq:diracBracketHydrodynamicalFieldFullPhaseSpace}), reduces to
\begin{equation}
  \left\{ f,g \right\}_D=\left\{ f,g \right\}_{Q_1,P_1},
  \label{}
\end{equation}
namely, the Poisson bracket on the reduced phase space.
Implementing the constraints (\ref{eq:constraintEquationsPhaseSpaceHydrodynamicCanonicalFormalism}) directly in transformation (\ref{eq:canonicalTransformationHydrodynamicalFieldReduction}), gives $Q_1=\rho$ and $P_1=\theta$, so that the Dirac bracket reduces to the Poisson bracket from Eq. (\ref{eq:poissonBracketReducedPhaseSpace}), where the reduced phase space of the system comprises the single pair of conjugate variables $\rho$ and $\theta$.
Hence from Eq. (\ref{eq:timeDerivativeDiracBracket}), it follows that the time-evolution of the conjugate pair of variables, is governed by the canonical field equations
\begin{align}
  \dot{\rho}\left( \v{r} \right)&=\left\{ \rho\left( \v{r} \right),H \right\}_D=\frac{\delta H}{\delta\theta\left( \v{r} \right)}, \label{eq:canonicalHydrodynamicalFieldEquations1} \\
  \dot{\theta}\left( \v{r} \right)&=\left\{ \theta\left( \v{r} \right),H \right\}_D=-\frac{\delta H}{\delta\rho\left( \v{r} \right)}.
  \label{eq:canonicalHydrodynamicalFieldEquations2}
\end{align}
The canonical Hamiltonian density (\ref{eq:Hamiltonian_density_GP}) becomes a strong equality on the reduced phase space, with
\begin{equation}
  \mathcal{H}=\rho\left[ \frac{\left( \grad{\theta} \right)^2}{2m}+\frac{g}{2}\rho +V\right]+\frac{\hbar^2}{8m\rho}\left( \grad{\rho} \right)^2.
  \label{eq:canonicalHamiltonianDensityHydrodynamicGPfield}
\end{equation}
The wave equations generated by the canonical field Eqs. (\ref{eq:canonicalHydrodynamicalFieldEquations1}) and (\ref{eq:canonicalHydrodynamicalFieldEquations2}) are, respectively, Eqs. (\ref{eq:continuityGPfield}) and (\ref{eq:QHJEGPfield}).
\subsubsection{\label{sec:GPfieldHydrodynamicCanonicalFormalismFJmethod}The Faddeev-Jackiw method}
When the Lagrangian is at most first order in the time derivatives of the fields, the Faddeev-Jackiw method provides a more direct approach to the Dirac-Bergmann algorithm.
In fact, the method was designed specifically for such Lagrangians.
Here we follow closely Jackiw's paper ``Quantization without tears'' \cite{jackiw1994quantization}, which we tailor to the specific system of interest.
The essence of the Faddeev-Jackiw approach, follows from the observation that the canonical momenta may be viewed as additional positional variables subject to their own Euler-Lagrange equations. 
In other words, the canonical equations of motion are equivalent to two Euler-Lagrange equations \cite{lanczos2012variational}:
\begin{align}
  \pd{L}{q_i}-\frac{d}{dt}\left( \pd{L}{\dot{q}_i} \right)&=-\pd{H}{q_i}-\dot{p}_i=0, \\
  \pd{L}{p_i}-\frac{d}{dt}\left( \pd{L}{\dot{p}_i} \right)&=\dot{q}_i -\pd{H}{p_i}=0. 
  \label{}
\end{align}
Therefore, given a Hamiltonian description governed by $H\left( q,p \right)$, it is always possible to construct a first order Lagrangian
\begin{equation}
  L\left( q,p \right)=\sum_i p_i\dot{q}_i - H\left( q,p \right),
  \label{}
\end{equation}
whose configuration space is identical to the Hamiltonian phase space \cite{jackiw1994quantization}, where the Euler-Lagrange equations for the Lagrangian $L\left( q,p \right)$ coincide with the canonical equations associated with $H\left( q,p \right)$.
Hence, if the Lagrangian presents itself in a first order form, one can readily identify the conjugate pairs of variables from the linear form in the velocities.
Let us examine how this works in practice for the present problem.\\
Consider the situation where the functions $\mathcal{A}_i$ from Eq. (\ref{eq:Lagrangian_density_first_order_in_velocities}) are linear in the fields, such that
\begin{equation}
  \mathcal{A}_i=\frac{1}{2}\sum_j\phi_j\omega_{ji},
  \label{eq:linearFieldFunctionsVelocityFactor}
\end{equation}
where $\omega$ is a matrix of constant coefficients.
Notice that the Lagrangian density (\ref{eq:Lagrangian_GP_polar}) of the Gross-Pitaevskii field appears in this form.
In such instances, the Lagrangian density may be presented as
\begin{equation}
  \mathcal{L}=\sum_{i,j}\frac{1}{2}\phi_j\omega_{ji}\dot{\phi}_i-\mathcal{H}.
  \label{eq:lagrangianDensityLinearFormLinearInFields}
\end{equation}
As a further observation, we note that the symmetric part of $\omega$ is equivalent to a total time derivative in $\mathcal{L}$ and, therefore, may be discarded \cite{jackiw1994quantization}.
Indeed, $\sum_{i,j}\phi_j\omega_{ij}\dot{\phi}_i=\partial_t\sum_{i,j}\phi_j\omega_{ij}\phi_i/2$ when $\omega_{ij}=\omega_{ji}$.
Hence, only the antisymmetric part of $\omega$ should be retained.
Taking this into account, the Euler-Lagrange equations for a field described by a Lagrangian density of the form (\ref{eq:lagrangianDensityLinearFormLinearInFields}), are equivalent to the following system of equations:
\begin{equation}
  \dot{\phi}_j=\sum_k \omega_{jk}^{-1}\frac{\delta H}{\delta\phi_k}.
  \label{eq:canonicalSystemEquationsLinearFormLinearFields}
\end{equation}
Thus, given the canonical Hamiltonian of a particular first order Lagrangian with linear functions $\mathcal{A}_i$, the associated canonical field equations may be obtained simply by evaluating the inverse of the antisymmetric part of the constant matrix $\omega$ characterising the system.
For the particular case of the Gross-Pitaevskii field described by the hydrodynamical Lagrangian density (\ref{eq:Lagrangian_GP_polar}), let us denote the field components by $\phi_1=\rho$ and $\phi_2=\theta$.
Then, in accordance with Eq. (\ref{eq:linearFieldFunctionsVelocityFactor}), it is clear that $\omega$ should solve the system of equations
\begin{align}
  \mathcal{A}_1&=\frac{1}{2}\left( \rho\omega_{11}+\theta\omega_{21} \right)=0 \\
  \mathcal{A}_2&=\frac{1}{2}\left( \rho\omega_{12}+\theta\omega_{22} \right)=-\rho,
  \label{}
\end{align}
so that
\begin{equation}
  \omega=
  \begin{pmatrix}
    0&-2\\
    0&0\\
  \end{pmatrix}
  .
  \label{}
\end{equation}
The antisymmetric part of this matrix, is
\begin{equation}
  \omega^A=
  \begin{pmatrix}
   0&-1\\
   1&0\\
  \end{pmatrix}
  ,
  \label{eq:omegaFJhydrodynamicalCanonicalFormalism}
\end{equation}
and should be identified with $\omega$ appearing in Eq. (\ref{eq:canonicalSystemEquationsLinearFormLinearFields}).
Notice that dropping the symmetric part of $\omega$ is equivalent to performing transformation (\ref{eq:LagrangianDensityTransformationTotalTimeDerivative}) on the Lagrangian density.
From the inverse of Eq. (\ref{eq:omegaFJhydrodynamicalCanonicalFormalism}), we find that the canonical equations (\ref{eq:canonicalSystemEquationsLinearFormLinearFields}) are equivalent to those of Eqs. (\ref{eq:canonicalHydrodynamicalFieldEquationRhoDot}) and (\ref{eq:canonicalHydrodynamicalFieldEquationThetaDot}).
For completeness, we note that Poisson brackets are defined so as to reproduce the canonical equations through Poisson commutation with the Hamiltonian, so that \cite{jackiw1994quantization}
\begin{equation}
  \left\{ f,g\right\}=\sum_{i,j}\int d^3\v{r}\frac{\delta f}{\delta\phi_i}\omega_{ij}^{-1}\frac{\delta g}{\delta\phi_j},
  \label{}
\end{equation}
which again yields expression (\ref{eq:poissonBracketReducedPhaseSpace}) as the Poisson bracket.
\section{Concluding remarks}
We have derived a hydrodynamic canonical formalism for the Gross-Pitaevskii field and eliminated redundant variables.
The physically relevant conjugate variables are the density and the phase of the condensate, which form a conjugate pair.
These results were obtained using both the Dirac-Bergmann and Faddeev-Jackiw methods.
The Faddeev-Jackiw method is a more direct approach, which lays out the problem in a much simpler form.
For instance, in addition to being more computationally demanding, the Dirac-Bergmann method relied on the addition of a suitably chosen total time derivative to the Lagrangian, in order to easily identify, at a later point, a canonical transformation which separates out the physical conjugate pair of variables from the pair of constraints.
In the Faddeev-Jackiw method, the same transformation arises by retaining the anti-symmetric part of $\omega$, which is assumed in the procedure.
Furthermore, no constraints appeared in the Faddeev-Jackiw method, highlighting the non-physical character of the constraints which appeared in the Dirac-Bergmann method.
\bibliographystyle{apsrev4-1}
\bibliography{References}
\end{document}